%% file: sn-article.tex
\documentclass[pdflatex,sn-mathphys-num]{sn-jnl}

\usepackage{graphicx}%
\usepackage{multirow}%
\usepackage{amsmath,amssymb,amsfonts}%
\usepackage{amsthm}%
\usepackage{mathrsfs}%
\usepackage[title]{appendix}%
\usepackage{xcolor}%
\usepackage{textcomp}%
\usepackage{manyfoot}%
\usepackage{booktabs}%
\usepackage{algorithm}%
\usepackage{algorithmicx}%
\usepackage{algpseudocode}%
\usepackage{listings}%
\usepackage{subcaption}
\usepackage{hyperref}
\usepackage{lineno}  

\theoremstyle{thmstyleone}%
\theoremstyle{thmstyletwo}%

\theoremstyle{thmstylethree}%

\begin{document}


\title[Article Title]{Towards Long-Range ENSO Prediction with an Explainable Deep Learning Model}


\author[1,2]{\fnm{Qi} \sur{Chen}}
\equalcont{These authors contributed equally to this work.}
\author[2]{\fnm{Yinghao} \sur{Cui}}
\equalcont{These authors contributed equally to this work.}
\author[3]{\fnm{Guobin} \sur{Hong}}
\author[4]{\fnm{Karumuri} \sur{Ashok}}
\author[5]{\fnm{Yuchun} \sur{Pu}}
\author[6,7]{\fnm{Xiaogu} \sur{Zheng}}
\author[8]{\fnm{Xuanze} \sur{Zhang}}
\author[2,9]{\fnm{Wei} \sur{Zhong}}
\author*[1]{\fnm{Peng} \sur{Zhan}}\email{zhanp@sustech.edu.cn}
\author*[2,9]{\fnm{Zhonglei} \sur{Wang}}\email{wangzl@xmu.edu.cn}

\affil[1]{\orgdiv{Department of Ocean Science and Engineering}, \orgname{Southern University of Science and Technology}, \orgaddress{\city{Shenzhen}, \postcode{518055}, \country{China}}}

\affil[2]{\orgdiv{Department of Statistics and Data Science}, \orgname{School of Economic, Xiamen University}, \orgaddress{\city{Xiamen}, \postcode{361005}, \country{China}}}

\affil[3]{\orgdiv{MOE Key Laboratory of Econometrics}, \orgname{Xiamen University}, \orgaddress{\city{Xiamen}, \postcode{361005}, \country{China}}}

\affil[4]{\orgdiv{Centre for Earth}, \orgname{Ocean and Atmospheric Sciences, University of Hyderabad}, \orgaddress{\city{Hyderabad}, \country{India}}}

\affil[5]{\orgdiv{Baidu Inc.}, \orgaddress{\city{Beijing}, \postcode{100085}, \country{China}}}

\affil[6]{\orgdiv{Shanghai Zhangjiang Institute of Mathematics}, \orgaddress{\city{Shanghai}, \postcode{201203}, \country{China}}}

\affil[7]{\orgdiv{International Global Change Institute}, \orgaddress{\city{Hamilton}, \country{New Zealand}}}

\affil[8]{\orgdiv{Key Laboratory of Water Cycle and Related Land Surface Processes}, \orgname{Institute of Geographic Sciences and Natural Resources Research, Chinese Academy of Sciences}, \orgaddress{\city{Beijing}, \postcode{100101}, \country{China}}}

\affil[9]{\orgdiv{Wang Yanan Institute for Studies in Economics}, \orgname{Xiamen University}, \orgaddress{\city{Xiamen}, \postcode{361005}, \country{China}}}

\input{main_parts/abstract}

\maketitle

\input{main_parts/1_introduction}

\input{main_parts/2_result}

\input{main_parts/3_conlusion_discussion}

\input{main_parts/4_materials_methods}

\input{main_parts/data_availability}


\bibliography{sn-bibliography}
\input{main_parts/contributions_interests}
\end{document}

%% file: main_parts/abstract.tex
\abstract{ El Niño-Southern Oscillation (ENSO) is a prominent mode of interannual climate variability with far-reaching global impacts. Its evolution is governed by intricate air-sea interactions, posing significant challenges for long-term prediction. In this study, we introduce CTEFNet, a multivariate deep learning model that synergizes convolutional neural networks and transformers to enhance ENSO forecasting. By integrating multiple oceanic and atmospheric predictors, CTEFNet extends the effective forecast lead time to 20 months while mitigating the impact of the spring predictability barrier, outperforming both dynamical models and state-of-the-art deep learning approaches. Furthermore, CTEFNet offers physically meaningful and statistically significant insights through gradient-based sensitivity analysis, revealing the key precursor signals that govern ENSO dynamics, which align with well-established theories and reveal new insights about inter-basin interactions among the Pacific, Atlantic, and Indian Oceans. The CTEFNet's superior predictive skill and interpretable sensitivity assessments underscore its potential for advancing climate prediction. Our findings highlight the importance of multivariate coupling in ENSO evolution and demonstrate the promise of deep learning in capturing complex climate dynamics with enhanced interpretability.}


%% file: main_parts/1_introduction.tex
\section{Introduction}\label{sec1}

El Niño-Southern Oscillation (ENSO) is one of the most prominent modes of interannual climate variability, characterized by shifts in sea surface temperatures (SST) across the tropical Pacific Ocean and the weakening of equatorial trade winds. ENSO exerts profound global influences on weather patterns, agriculture, and socio-economic systems by driving variability in precipitation, temperature, as well as extreme events such as droughts and floods \cite{doi:10.1126/science.1132588,ashok2009nino,timmermann2018nino}. Traditional statistical and dynamical models have demonstrated predictive skill within a lead time of about 12 months (where effective forecast lead time is defined as the period during which the correlation between the forecasted ENSO index and the observed value remains above 0.50)\cite{chenPredictabilityNino1482004, 10.1093/nsr/nwy105, capotondi2015optimal,penland1996stochastic,kondrashov2005hierarchy,lima2009statistical}. However, ENSO prediction remains a formidable challenge due to the system’s inherent nonlinearity, stochasticity, and multivariate dependencies \cite{latif1998review, caneExperimentalForecastsNino1986, doi:10.1126/science.1237554, 10.1093/nsr/nwy105, https://doi.org/10.1002/2017RG000568,zhang2022recent}, with one of the most persistent limitations being the spring predictability barrier (SPB) \cite{webster1992monsoon}.

Recent advances in deep learning (DL) have demonstrated transformative potential in ENSO forecasting. Convolutional neural networks (CNNs) have demonstrated remarkable skill in capturing spatial features, extending the effective forecast lead time beyond 15 months \cite{hamDeepLearningMultiyear2019,hamUnifiedDeepLearning2021}. Other DL techniques, such as recurrent neural networks (RNNs)\cite{mahesh2019forecasting}, convolutional long short-term memory (LSTM) neural networks\cite{gupta2020prediction}, graph neural networks\cite{muENSOGTCENSODeep2022}, and transformers \cite{zhouSelfattentionbasedNeuralNetwork2023,zhang2024transformer,muIncorporatingHeatBudget2024}, have further enhanced spatiotemporal dependency modeling, successfully extending forecast lead times to 18 months and beyond, with some models achieving performance exceeding 20 months\cite{muIncorporatingHeatBudget2024,huDeepResidualConvolutional2021,wangInterpretableDeepLearning2023,wangRoleSeaSurface2024}. Among these advancements, transformer-based architectures have emerged as a particularly powerful approach, leveraging self-attention mechanisms to capture complex, long-range dependencies across three-dimensional ENSO dynamics. However, despite these improvements, no single architecture is universally optimal for ENSO forecasting. For instance, CNNs and transformers, two of the most widely adopted DL models in this field, each exhibit distinct strengths and limitations. CNNs, while adept at extracting spatial features, struggle to capture long-term dependencies. Meanwhile, transformers, despite their ability to model global interactions, often require large datasets and lack the inductive biases necessary to recognize key local ENSO precursors \cite{hamDeepLearningMultiyear2019,dosovitskiy2021imageworth16x16words,mehta2022mobilevitlightweightgeneralpurposemobilefriendly}. To address these challenges, CNN-transformer hybrid models have been introduced in geoscience \cite{bai2022rainformer,li2023convtransnet,chen2024leformer}, and their application to ENSO forecasting has demonstrated robust performance up to 18 months across diverse test datasets\cite{lyuResoNetRobustExplainable2024a}. However, this current approach predominantly relies on an encoder-based, SST-only architecture, failing to incorporate the critical ocean-atmosphere interactions that govern ENSO evolution\cite{muENSOASC100ENSO2021}.

In response to these limitations, we introduce CTEFNet (Convolutional Transformer ENSO Forecast Network), a novel hybrid deep learning model that synergistically integrates CNNs within a transformer encoder-decoder architecture. By leveraging the complementary strengths of these architectures and incorporating a comprehensive set of oceanic and atmospheric variables, CTEFNet effectively captures the multivariate precursors of ENSO evolution. Our model achieves state-of-the-art forecast performance, extending the effective lead time to 20 months. Moreover, CTEFNet successfully mitigates the SPB, underscoring its robustness and reliability in long-range ENSO forecasting.

Beyond its predictive superiority, CTEFNet demonstrates physical interpretability through a novel gradient-based sensitivity analysis\cite{JSSv102i07, bano2025ai}, inspired by the principles of adjoint modeling techniques\cite{marotzke1999construction,losch2007adjoint,zhang2012sensitivity,zhanSensitivityStudiesRed2018c}. Unlike conventional sensitivity analysis that relies on an ensemble of forward modeling with perturbed inputs \cite{hamDeepLearningMultiyear2019,zhouSelfattentionbasedNeuralNetwork2023}, adjoint sensitivity analysis, widely used in ocean and climate modeling, quantifies how perturbations in an objective function propagate backward through the evolution of a system
\cite{zhang2012sensitivity}. However, adjoint models are computationally expensive and often constrained by linearity assumptions, which limit their applicability to complex, nonlinear systems. In contrast, backpropagation gradients offer a computationally efficient and nonlinear assessment of input influence on the objective, in this case, ENSO evolution. This approach enables a systematic evaluation of the relative importance of different inputs across varying temporal and spatial scales \cite{wangInterpretableDeepLearning2023, lyuResoNetRobustExplainable2024a, wangUnderstandingLowPredictability2024a}. While conventional DL gradient-based methods primarily evaluate how input perturbations affect predictive performance\cite{muIncorporatingHeatBudget2024}, our approach combines DL gradients with adjoint principles to derive a dynamic, spatiotemporally evolving sensitivity analysis, revealing the physical mechanisms that drive ENSO formation from a data-driven perspective.

Our sensitivity analysis uncovers physical precursors of ENSO events consistent with established mechanisms\cite{bjerknes1969atmospheric, jin1997equatorial, ham2013sea,park2022role,wang2006overlooked,wang2021joint,fan2024coupling,xie2009indian,hameed2018model}. Furthermore, it reveals new insights into the development of inter-basin interactions, advancing our understanding of ENSO’s global influence. By improving both predictive skill and interpretability, this study highlights the critical role of multivariate coupling in ENSO dynamics and underscores the value of deep learning in climate science. These findings establish CTEFNet as a practical and scalable solution for long-term ENSO forecasting, bridging the gap between data-driven predictions and physical understanding of climate variability.

%% file: main_parts/2_result.tex
\section{Results}

CTEFNet, built upon a novel CNN-transformer hybrid architecture (MATERIALS AND METHODS), integrates key ocean-atmosphere variables from the Coupled Model Intercomparison Project Phase 6 (CMIP6) SSP370 dataset, spanning 2015 to 2100. This dataset, representing a medium-to-high emissions scenario, incorporates future climate forcings, providing a comprehensive depiction of ENSO dynamics in a warming climate. Using a 12-month predictor window, CTEFNet incorporates SST, heat content (HC), mixed layer depth (MLD), sea surface salinity (SSS), sea level pressure (SLP), and the zonal and meridional components of ocean surface current velocity (UO, VO) and wind stress (TAUU, TAUV). CTEFNet predicts the evolution of the Niño 3.4 index over a 24-month horizon, and its performance is rigorously assessed through prediction skill evaluations and sensitivity analysis, utilizing reanalysis data from the Global Ocean Data Assimilation System (GODAS) and the fifth-generation ECMWF atmospheric reanalysis (ERA5) from 1980 to 2021.

\subsection{Predictive Skill of CTEFNet in ENSO Forecasting}

\begin{figure*}[htbp]
\centering
\begin{subfigure}[t]{1\linewidth}
		\centering
		\includegraphics[width=1\linewidth]{
                {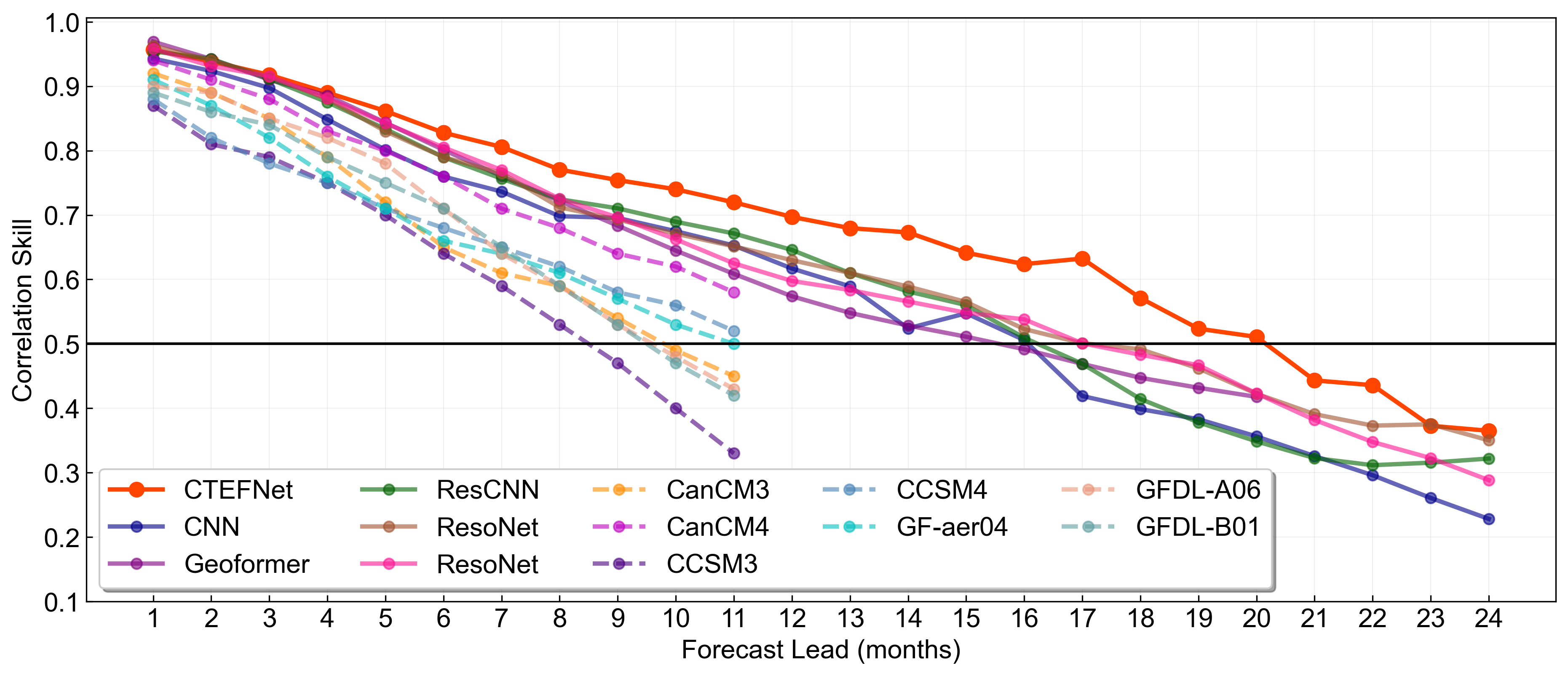}}            
\end{subfigure}

\caption{\textbf{ENSO correlation skill in CTEFNet and other models.}  The all-season  correlation skill of the three-month-moving-averaged Niño 3.4 index as a function of the forecast lead month in CTEFNet (solid orange), CNN(solid deep blue), Geoformer (solid purple), ResCNN (solid green), ResoNet (solid brown), STPNet (solid pink) and the dynamical forecast systems included in the NMME project (dash with other colors). The validation period is between 1980 and 2021. } 

\label{fig: result}

\end{figure*}

To assess the forecasting skill of CTEFNet, we utilize the Niño 3.4 index, a widely used SST anomaly-based metric, to characterize ENSO variability. CTEFNet exhibits superior forecast skill, significantly outperforming the North American Multi-Model Ensemble (NMME) dynamical models \cite{TheNorthAmericanMultimodelEnsemblePhase1SeasonaltoInterannualPredictionPhase2towardDevelopingIntraseasonalPrediction}, as well as state-of-the-art DL approaches including CNN  \cite{hamDeepLearningMultiyear2019}, Geoformer\cite{zhouSelfattentionbasedNeuralNetwork2023}, ResCNN\cite{huDeepResidualConvolutional2021}, ResoNet\cite{lyuResoNetRobustExplainable2024a}, and STPNet\cite{wangRoleSeaSurface2024}. As shown in Fig.~\ref{fig: result}, CTEFNet demonstrates markedly higher predictive accuracy in terms of correlation coefficient, particularly for mid-to-long-term forecasts with lead times beyond 6 months. It maintains all-season correlation skills above 0.7 for forecast leads up to 12 months, extending its predictive horizon by 3 months longer than existed deep learning models, and by over 5 months longer than dynamical models. Additionally, CTEFNet sustains all-season correlation skills above 0.6 for forecast leads extending beyond 17 months, a 4-month improvement over existing deep learning models. Although previously developed DL models exhibit strong forecasting performance on specific datasets, their forecasting capabilities typically diminish when trained on the CMIP6 SSP370 data, often limiting their effective lead time to around 17 months. This limitation may stem from these models' insufficient physical interpretability, potentially reducing their ability to generalize robustly across datasets characterized by diverse physical mechanisms.

\begin{figure*}[h]
\centering

\begin{subfigure}[t]{0.9\linewidth}
    \centering
    \captionsetup{justification=raggedright, singlelinecheck=false, position=top} 
    \includegraphics[width=1\linewidth]{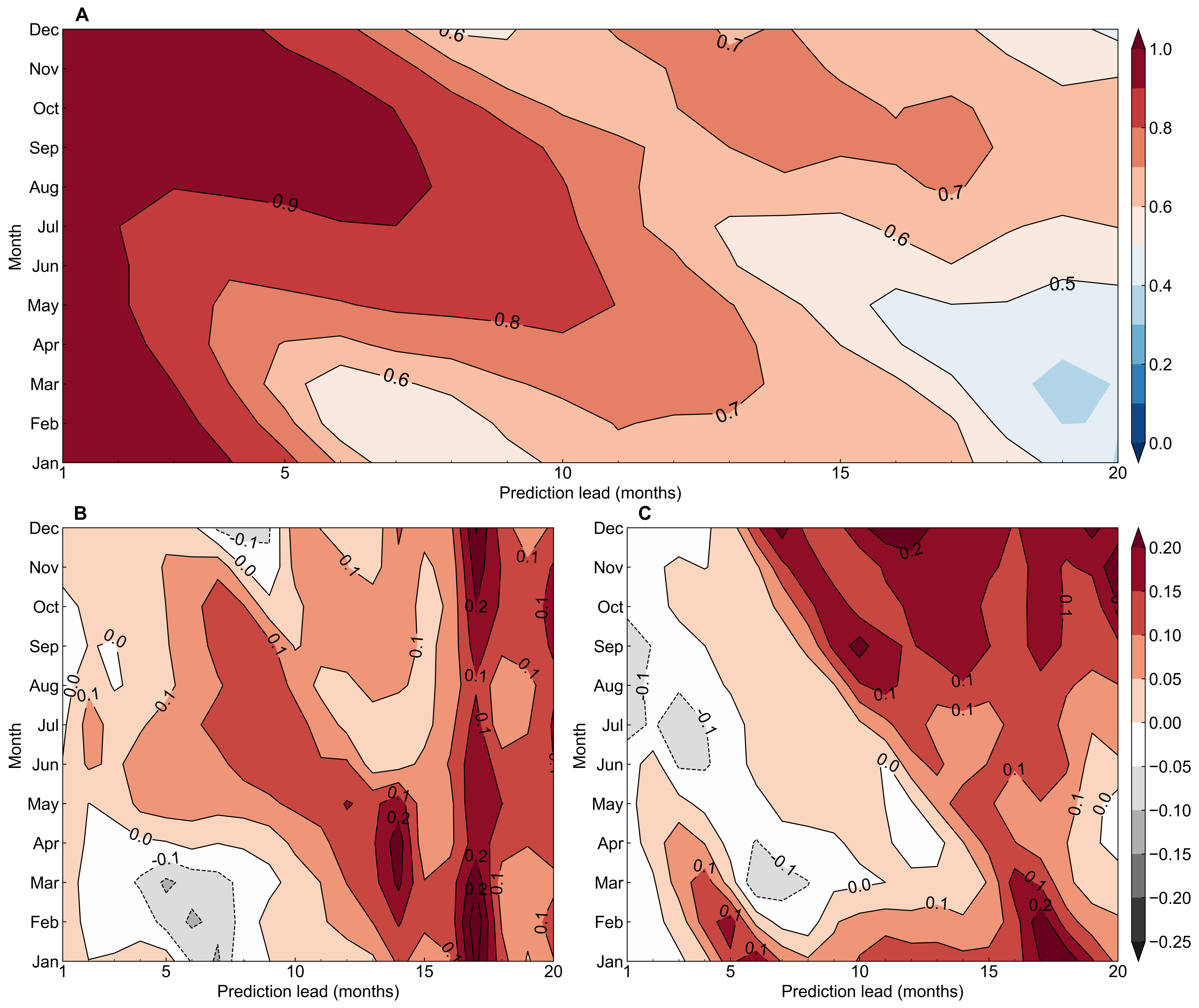}
  
\end{subfigure}
\caption{\textbf{The seasonality and lead-time of CTEFNet's performance.} \textbf{(A)} Contour plot of correlation skills for CTEFNet across calendar months during the test period (1980–2021) at different lead times. The horizontal axis denotes the forecast lead month, while the vertical axis represents the calendar month. \textbf{(B)} Same as A, but showing the correlation skill difference between CTEFNet and CNN. \textbf{(C)} Same as (A), but showing the correlation skill difference between CTEFNet and Geoformer.}
 \label{fig: seasonal_skill}
\end{figure*}

To further evaluate seasonal variations in forecast performance, Fig. \ref{fig: seasonal_skill} A presents the seasonal correlations skills of our CTEFNet across different lead times. The correlation skill remains above 0.5 for over 20 months from June to December, and for 16 months during the boreal spring (March to May), despite the visible influence of the SPB. Additionally, Fig. \ref{fig: seasonal_skill} B and C show that CTEFNet significantly outperforms CNN and Geoformer in predicting both autumn and winter conditions, while also excelling in mid- to long-term predictions for the boreal spring. These results highlight CTEFNet's capability not only in achieving high accuracy but also in mitigating the challenges of seasonal forecast degradation.

\subsection{Unveiling ENSO Precursors through Gradient-Based Sensitivity Analysis}

\begin{figure*}[p]
\centering

\begin{subfigure}[t]{0.95\linewidth}
    \centering
    \captionsetup{justification=raggedright, singlelinecheck=false, position=top} 
    \includegraphics[width=1\linewidth]{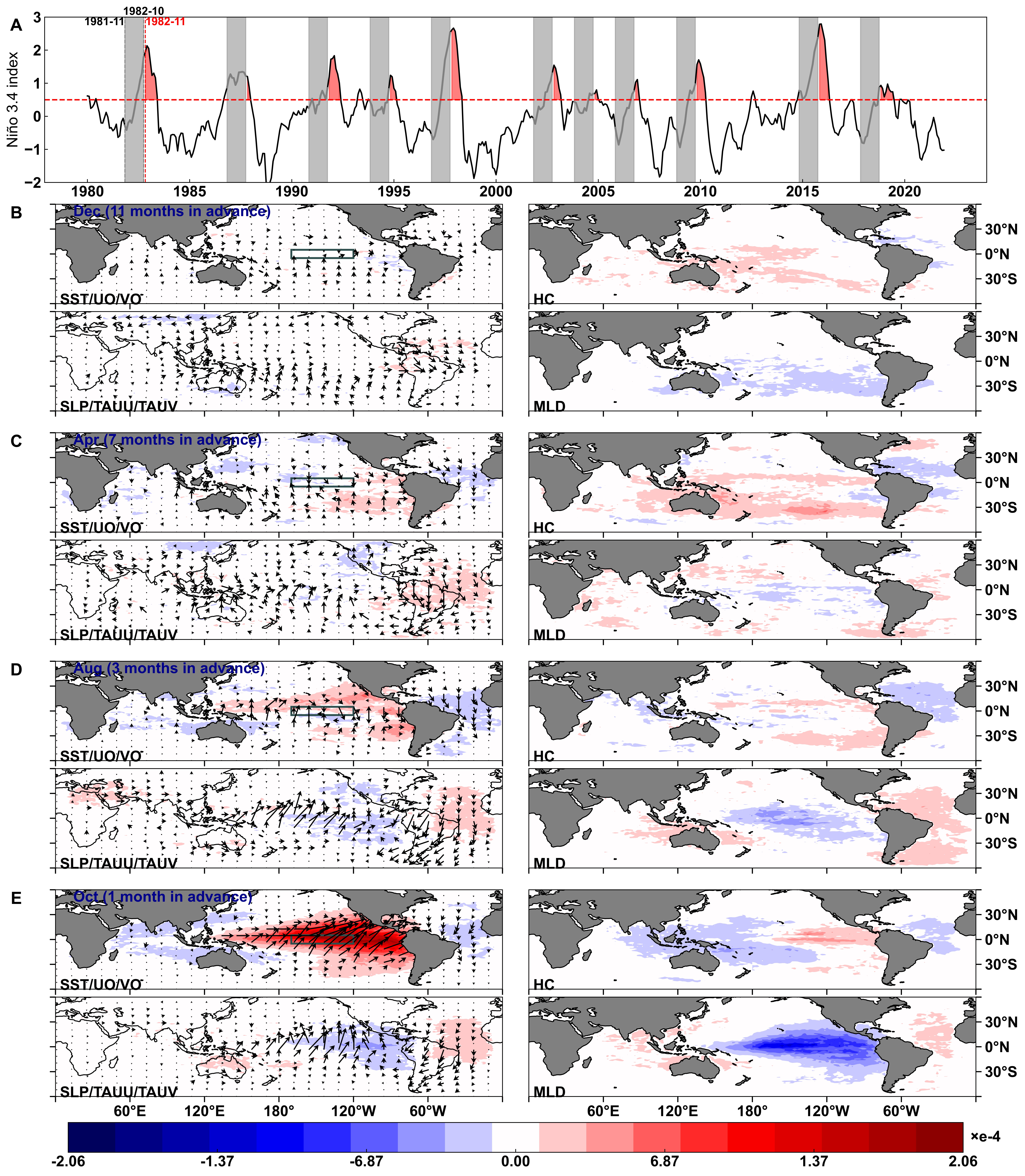}
\end{subfigure}

\caption{\textbf{The precursors and underlying mechanisms of ENSO forecasting revealed by CTEFNet.} (\textbf{A})  Sensitivity analysis periods, with red indicating predicted target periods and gray representing 12-month input periods.
(\textbf{B} to \textbf{E})  Averaged sensitivities across multiple El Niño events, retaining only grid point values that are statistically significant at the 95\% confidence level, illustrating the contributions of various predictors across different months. Colors denote the sensitivities of SST, HC, SLP, and MLD, while vectors represent those of UO, VO, TAUU, and TAUV. The green box marks the Niño 3.4 region.}

 \label{fig: sensitivity}
\end{figure*}

By computing the backpropagation gradients of the multivariate inputs with respect to the Nin\~{n}o 3.4 index, we quantify the relative influence of key ocean-atmosphere variables across different lead times, a measure we refer to as sensitivity (Fig. \ref{fig: sensitivity}, MATERIALS AND METHODS, and Fig. S1-S5). Through systematic sensitivity analysis based on this novel approach, we demonstrate that CTEFNet captures the seasonal evolution and propagation of ENSO signals prior to its maturity. This capability, underpinned by interpretable realistic representation of ENSO's physical mechanisms, significantly contributes to the model's enhanced predictive performance. 

The sensitivity analysis was conducted for 11 El Niño events (1982, 1987, 1991, 1994, 1997, 2002, 2004, 2006, 2009, 2015, and 2018) from 1980 to 2021. The prediction targets were determined based on the Niño 3.4 index in November of the El Niño year (when El Niño typically peaks) and the subsequent months (when the index exceeds 0.5), with corresponding inputs derived from the 12-month period preceding November  (Fig. \ref{fig: sensitivity} A). To identify robust precursor signals, we computed the average sensitivities across these El Niño events at each lead month, revealing a sequence of physically meaningful precursor patterns. These sensitivities were statistically assessed using a Student’s t-test, and only grid points with sensitivities significantly different from zero at the 95\% confidence level are retained. This filtering step ensures that our interpretations are grounded in statistically significant and robust signals, thereby enhancing the reliability of the identified precursors (Fig. \ref{fig: sensitivity} B–E). This approach enables a systematic quantification of both positive and negative sensitivities, providing clear insights into the distinct impact of each variable on ENSO evolution. Additionally, the use of normalized input data allows for direct comparison of variable contributions across different time periods and regions, establishing a comprehensive framework for understanding ENSO predictability. To provide a broader perspective, we also conducted a parallel sensitivity analysis for La Niña events (Fig. S5). While La Niña sensitivity patterns generally exhibit a spatial distribution opposite to that of El Niño events, notable asymmetries emerge in the seasonal evolution of inter-basin interactions across the equatorial oceans. However, as our primary focus is El Niño sensitivity, a detailed analysis of La Niña falls beyond the scope of this study. Nevertheless, these findings highlight the potential for further investigations into asymmetric ENSO dynamics, which could refine our understanding of ENSO predictability.

Our sensitivity analysis reveals that CTEFNet captures the early precursors of El Ni\~{n}os, particularly through the sensitivity fields of HC and MLD in the equatorial Pacific observed in November, approximately 11 months before the El Ni\~{n}o peaks (Fig. \ref{fig: sensitivity} B). These signals align with the recharge phase of the recharge oscillation mechanism \cite{jin1997equatorial}, where positive HC sensitivity in the equatorial western Pacific (WP) indicates the accumulation of warm surface waters and a deepening thermocline, establishing favorable preconditions for El Ni\~{n}o development. Concurrently, negative MLD sensitivity across the tropical Pacific, typically associated with a thickened ocean barrier layer\cite{sprintall1992evidence, maes2002salinity}, suggests a suppression of vertical entrainment and mixing, facilitating heat retention at the surface and reinforcing conditions conducive to El Ni\~{n}o initiation\cite{maes2005importance, rudzin2018influence}.

By April (eight months before El Ni\~{n}o peaks), the El Ni\~{n}o-related signals are ubiquitous throughout the global equatorial oceans. (Fig. \ref{fig: sensitivity} C). The positive sensitivity of HC in the WP intensifies and extends into the central Pacific (CP), eastern Pacific (EP), and eastern Indian Ocean (IO). This eastward expansion of the Pacific warm pool coincides with a strengthening positive SST sensitivity in the tropical EP and an intensified westerly wind sensitivity over the tropical Pacific, indicating multivariate ocean-atmosphere interactions preceding El Ni\~{n}o events. These patterns mark the initiation of the Bjerknes positive feedback mechanism\cite{bjerknes1969atmospheric}, wherein a rise in SST in the EP weakens the zonal temperature gradient, enhancing westerly wind, which in turn amplifies eastward warm water transport, reinforcing positive SST anomalies in the EP. In contrast, HC and SST in the North Tropical Atlantic (NTA) exhibit negative sensitivities (Fig. \ref{fig: sensitivity} C). Cooling in the NTA has been shown to suppress convection in the tropical Atlantic\cite{ham2013sea}, which alters the position and intensity of the Pacific Intertropical Convergence Zone (ITCZ) through air-sea interactions, particularly via moist static energy feedback processes\cite{alexander2002influence, wuAtmosphericDynamicThermodynamic2017, park2022role}. These atmospheric adjustments weaken the Pacific trade winds, reduce upwelling, and facilitate the accumulation of warm water in the EP and CP. This sequence of processes ultimately enhances the eastward propagation of warm Kelvin waves, disrupts the normal Pacific circulation, and establishes a critical inter-basin teleconnection pathway linking Atlantic variability to ENSO\cite{yu2016effects, jiang2021impacts}. Additionally, increased ocean current speeds and enhanced wind stress to the south in the tropical Atlantic contribute positively to El Ni\~{n}o formation in our sensitivity analysis. These changes could intensify cross-equatorial heat transport, modulate surface fluxes, and alter large-scale atmospheric circulation, which may promote adjustments in the Walker circulation and further amplifying El Ni\~{n}o development.

From April to August (three months before El Ni\~{n}o peaks), the positive SST sensitivity field in the EP progressively expands westward into the CP and WP, while westerly wind stress sensitivity in the WP intensifies and negative MLD sensitivity in the EP and CP becomes more pronounced (Fig.~\ref{fig: sensitivity} D). These shifts align with the canonical evolution of El Niño, highlighting the complex multivariate interactions driving ENSO development.  The evolving SST pattern alters atmospheric pressure systems, further weakening the trade winds, which in turn amplifies warm SST anomalies across a broader region of the Pacific. As the trade winds weaken and SST variations intensify, shoaling MLD in the EP and CP reduces heat exchange between the deep ocean and surface waters, accelerating the warming of surface waters. During this period, the negative SST sensitivity field in the NTA propagates southward, encompassing the entire tropical Atlantic. This cooling suppresses convection and strengthens descending motions over the Atlantic, reinforcing the descending branch of the Walker circulation and leading to compensatory westerly wind anomalies over the tropical Pacific\cite{wang2021joint}. Additionally, wind stress sensitivities associated with this Atlantic cooling, characterized by easterlies off Central America and over South America, further strengthen the teleconnection between the Atlantic and Pacific, establishing a robust link between Atlantic variability and ENSO dynamics\cite{wang2021joint, wang2006overlooked}. Simultaneously, negative SST sensitivities emerge in the eastern IO, influencing the intensity and spatial configuration of the Indian Ocean Dipole (IOD). This shift in the IOD modulates the Walker circulation, generating compensatory westerly wind anomalies over the tropical Pacific, further promoting El Ni\~{n}o development \cite{hameed2018model, fan2024coupling}. The interaction between the IO and Pacific SSTs further strengthens the ocean-atmosphere feedback mechanisms driving the intensification of El Ni\~{n}o events. 

By October (one month before El Ni\~{n}o peaks), SST sensitivity in the tropical Pacific continues to increase as El Ni\~{n}o conditions mature (Fig. \ref{fig: sensitivity} E), further amplifying ENSO’s ocean-atmosphere feedback loop. Simultaneously, wind stress sensitivity and ocean surface current sensitivity indicate that strengthening westerly wind stress and intensified westward ocean current velocity transport warm water from the WP to the CP and EP, leading to warm water accumulation in the EP and a corresponding decrease in HC in the WP. This results in contrasting HC sensitivities: while the WP experiences a negative impact, the EP exhibits a positive response. Moreover, negative MLD sensitivivty field across the tropical pacific suggests that shoaling of mixed layer further enhances surface heat accumulation, reinforcing the positive feedback loop driving El Ni\~{n}o development. Concurrently, negative SST sensitivity persists in the tropical Atlantic, along with basin-wide negative sensitivity fields over the northern and equatorial IO, the South China Sea, and off the northern Australian coasts. Notably, cooling in the IO induces divergence and westerly winds over the western tropical Pacific Ocean, a process consistent with the Matsuno–Gill response \cite{matsuno1966quasi, gill1980some}. This dynamic adjustment triggers the eastward propagation of downwelling Kelvin waves, which further reinforces El Ni\~{n}o development\cite{xie2009indian}. The resulting SST sensitivity patterns are accompanied by notable negative SLP sensitivity fields in the EP and CP, alongside positive SLP sensitivity fields over the Atlantic and Australian regions. This pressure pattern modulates large-scale wind fields and strengthens ocean current sensitivity toward the eastern Pacific, further amplifying the transport of warm water to the CP and EP, reinforcing El Ni\~{n}o intensification.

The identified precursor signals and sensitivity patterns derived from CTEFNet align closely with the canonical evolution of El Niño, demonstrating that our model effectively represents key physical processes underpinning ENSO development and propagation.

%% file: main_parts/3_conlusion_discussion.tex
\section{Conclusion and Discussion}

Recent advancements in DL have revolutionized ENSO forecasting, offering a powerful data-driven framework capable of capturing its highly nonlinear dynamics. However, despite these significant strides, challenges persist in achieving robust and interpretable multivariate ENSO predictions. To address these gaps, we introduce CTEFNet, a novel CNN-transformer hybrid model designed to effectively capture the coupled spatiotemporal interactions governing ENSO evolution. Compared to the CNN-transformer model proposed by Lyu et al. \cite{lyuResoNetRobustExplainable2024a}, which relies on an encoder-based architecture to process only SST through separate CNN and Transformer modules, CTEFNet employs a distinctive encoder-decoder framework. A key advantage of this design lies in its sequential prediction capability, which enables CTEFNet to dynamically capture the evolving multivariate interactions at each time step. This architecture makes CTEFNet particularly well-suited for sequence generation and long-range forecasting. CTEFNet’s predictive performance is further enhanced through training on an ensemble dataset from CMIP6, allowing it to account for subtle differences in physical mechanisms across multiple climate models. This ensemble-based learning approach significantly improves CTEFNet’s ability to capture implicit multivariate ENSO dynamics, which may not be fully represented in a deterministic dynamical model.  As a result, CTEFNet achieves effective long-lead ENSO forecasts up to 20 months while significantly reducing the SPB, outperforming both traditional dynamical models and state-of-the-art DL models. 

Beyond its superior predictive performance, CTEFNet stands out for its high interpretability, addressing a longstanding limitation of DL-based climate models. We have developed a gradient-based sensitivity analysis methodology, envisioned as a practical alternative to conventional adjoint models, to identify global ENSO precursors and their underlying mechanisms. Unlike traditional adjoint methods, which are often computationally prohibitive and constrained by linearity assumptions, our gradient-based approach offers a more flexible and nonlinear representation of ENSO dynamics. This method overcomes the common challenge of inadequately managing nonlinear responses in physical oceanography and climate science, providing a more faithful representation of complex system dynamics. Meanwhile, a major strength of our approach is its efficiency, derived naturally as a byproduct of the DL model, eliminating the need for additional computationally intensive integration steps. This inherent efficiency makes it highly scalable and well-suited for large-scale climate simulations. The integration of this methodology represents a step forward in enhancing the robustness and applicability of climate sensitivity analysis, particularly in scenarios where nonlinear interactions are significant. Our sensitivity analysis with CTEFNet reveals physical precursors to El Niño events that are consistent with some established mechanisms. Specifically, the sequential buildup of heat in the WP and its eventual release to the CP and EP aligns with the recharge-oscillation mechanism. Additionally, the positive feedback loop in the tropical Pacific, characterized by rising SSTs in the CP and EP and the amplification of westerly winds, strongly reflects the Bjerknes feedback mechanism. Notably, our analysis also highlights the role of inter-basin interactions in ENSO variability, revealing that cooling in the tropical Atlantic and Indian Oceans influences large-scale wind patterns and atmospheric circulation, ultimately modulating El Niño formation through cross-basin teleconnections.

In this regard, another recent approach using Swin Transformer \cite{Liu_2021_ICCV} could achieve comparable predictive performance of Niño 3.4 index with good computational efficiency (Fig. S6 A), yet it seems not as robust as CTEFNet for failing to provide physically meaningful interpretability (Fig. S6 B, C). This limitation likely stems from the absence of strong inductive biases, which are inherently provided by the CNN component in CTEFNet. CNNs enable robust feature extraction, maintaining stability even in the presence of noisy data \cite{zeiler2014visualizing, cao2022random, wang2024theoretical}. While Swin Transformer incorporates spatial localization through its hierarchical structure and sliding windows, its inductive biases are weaker and less explicitly defined compared to those of CNNs. Consequently, Swin Transformer may exhibit instability when trained on limited datasets, leading to gradient fluctuations and reduced robustness in sensitivity analysis \cite{xu2021vitae, 10186881}.

Furthermore, CTEFNet’s sensitivity analysis reveals new insights into El Niño’s seasonal evolution, particularly regarding inter-basin influences. For instance, we identify a persistent cooling signal in the tropical Atlantic from spring through autumn, with easterly wind anomalies near Central and South America strengthening the teleconnection between the Atlantic and Pacific Oceans. In the Indian Ocean, sensitivity fields extend from the eastern basin in summer to the northern and equatorial regions by autumn, underscoring the evolving nature of inter-basin interactions. These findings paint a more dynamic picture of ENSO’s seasonal progression than conventional models suggest, indicating that inter-basin interactions may be far more critical in driving ENSO’s peak-phase characteristics than previously recognized.

Despite its strengths, CTEFNet’s current implementation remains focused on predicting the Niño 3.4 index, limiting its direct application to ENSO diversity\cite{ashok2007nino, capotondi2013enso, marathe2015revisiting}.  A natural extension of this work would involve enhancing the model’s capability to differentiate between Modoki and canonical ENSO events, which exhibit distinct climatic impacts. Additionally, while CTEFNet has demonstrated significant improvements in mitigating the SPB, it remains a formidable challenge. The persistent decline in model performance during spring is typically attributed to the complex dynamics of the tropical climate system, including shifts in wind patterns and ocean currents that are not well-captured by existing models. Addressing this limitation requires a concerted effort to dissect the underlying mechanisms of SPB and to develop refined modeling approaches that can account for these intricate seasonal variations. Innovative methodologies, possibly integrating higher-resolution data and advanced machine learning techniques, could improve predictions during this challenging season. Such advancements would not only enhance the accuracy of climate models like CTEFNet but also broaden their applicability in real-world climate strategy and policy-making, where understanding and anticipating climate variability is crucial.

%% file: main_parts/4_materials_methods.tex
\section{Materials and Methods}

\subsection{Data and processing methods}

The performance of DL models is largely determined by both the quantity and quality of training data. However, the observation data for extreme climate events, such as ENSO, is often insufficient to provide adequate sampling. To address this limitation, simulation data from 18 CMIP6 climate models (2015–2100) are utilized for model training (Table S1). For model evaluation and selection, reanalysis datasets from the Ocean Reanalysis System 5 (ORAS5) and ERA5 (1958–1978) are used as validation sets. To further evaluate the model's generalization ability, it is tested using data from GODAS and ERA5 (1980–2021).

Before inputting the data into CTEFNet, a uniform preprocessing procedure is applied. First, monthly anomalies for each input variable are calculated by removing long-term trends and climatology, and  this operation is performed separately for each CMIP6 model. The data are then standardized to a uniform spatial resolution of $1^\circ \times 2^\circ$ through linear interpolation, covering the spatial domain from $60^\circ \text{S}$ to $60^\circ \text{N}$ in latitude and $0^\circ$ to $360^\circ$ in longitude. Grids corresponding to land areas (except for wind stress and sea level pressure) and missing data are assigned a value of zero. The processed fields are then normalized and concatenated along the layer axis to form datasets comprising nine layers.

The input data includes SST, HC, MLD, SSS, SLP, UO, VO, TAUU, and TAUV from the current and previous eleven months. These variables are combined in an overlapping manner, resulting in a data format of size $[12 \times 9 \times 120 \times 180]$, where the four dimensions represent the temporal duration of the input data, the number of variable types, and the latitude and longitude grids. The target variable for training CTEFNet is the Niño 3.4 index for the subsequent 24 months, with the corresponding data format being $[24]$.

\subsection{Architecture of CTEFNet}

\begin{figure*}[t]
\centering

    \centering
    \includegraphics[width=1\linewidth]{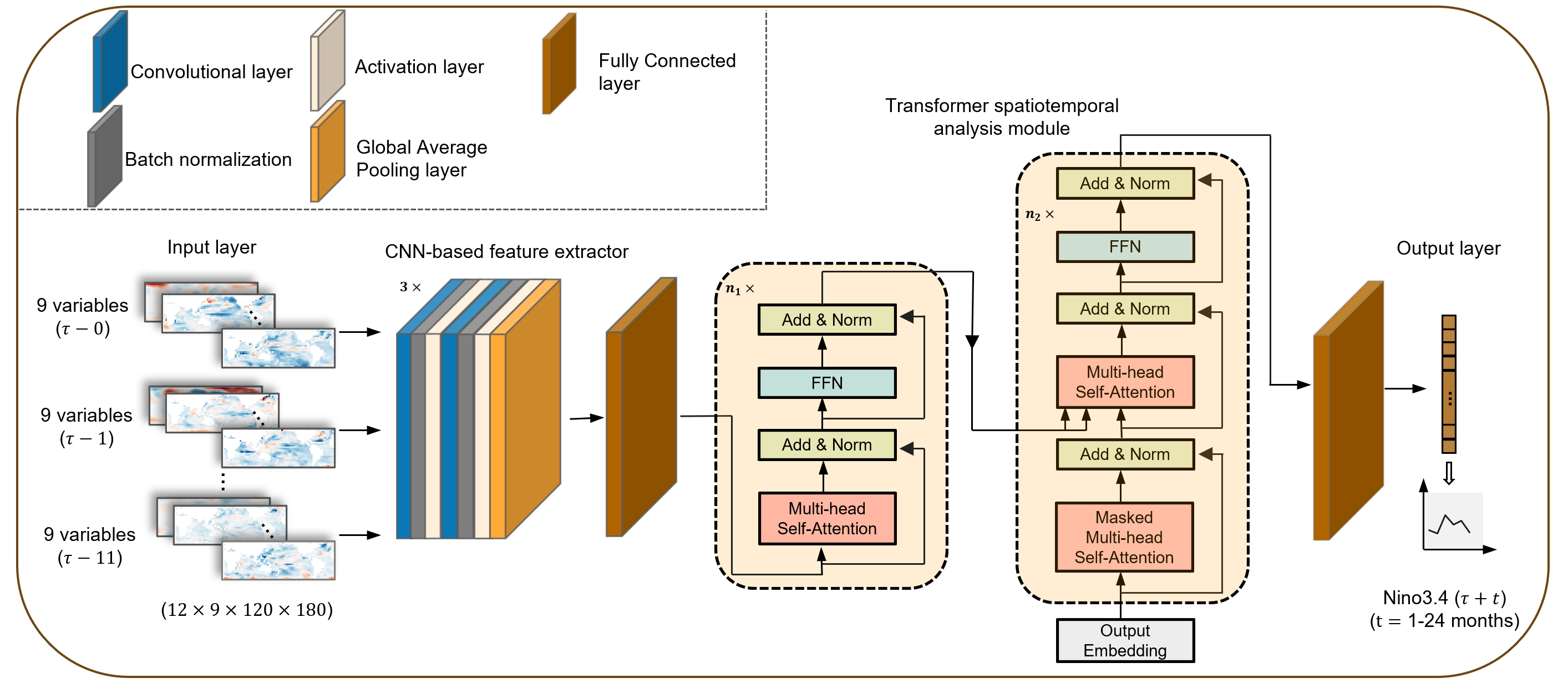}
    

\caption{\textbf{Architecture of CTEFNet for ENSO predictions.} CTEFNet consists of an input layer, a CNN-based feature extractor, a Transformer spatiotemporal analysis module, two fully connected layers, and an output layer. The input predictors include SST, HC, MLD, SSS, SLP, UO, VO, TAUU, and TAUV anomaly fields, all spanning 12 consecutive months in the region defined by ($60^{\circ}S-60^{\circ}N$, $0^\circ-360^\circ$E). The Niño 3.4 index for the subsequent 24 months serves as the predictands for supervised training.}

\label{fig: model}
\end{figure*}

CTEFNet comprises two primary components: a CNN-based feature extractor and a Transformer spatiotemporal analysis module (Fig. \ref{fig: model}). The CNN-based feature extractor performs multi-scale downsampling of input variables, capturing key regional spatial features. The Transformer module utilizes self-attention mechanisms and parallel processing to model multivariable relationships and long-range dependencies in sequential data. Unlike previous transformer-based models, which predict the entire forecast region \cite{zhouSelfattentionbasedNeuralNetwork2023,zhang2024transformer}, CTEFNet directly predicts the Niño 3.4 index. This design enables early downsampling within the pipeline, optimizing computational efficiency. As a result, CTEFNet can process larger, global input data, improving ENSO prediction while enhancing sensitivity analysis of global multivariate patterns.

To extract the spatiotemporal features from the input variables, we employ a stack of three CNN-based blocks. Each block consists of two convolution layers, two batch normalization layers, two ReLU activation functions, and one global average pooling layer. The convolution operations, with their local receptive fields, enable the model to capture critical local information while minimizing global context noise. The hierarchical structure of the CNN blocks also facilitates the extraction of multi-scale spatial features, making them well-suited for focusing on specific regions. The Transformer, with its encoder-decoder architecture, excels in modeling spatiotemporal sequences. Its self-attention mechanism efficiently captures long-range dependencies across time steps and spatial locations, enabling the model to identify and leverage key factors that drive climate change, regardless of their position in the sequence.

\subsection{Model training strategy}

CTEFNet processes batches of input variables (batch size = 8), where each batch contains 12 consecutive months of data as predictors, and the Niño 3.4 index for the subsequent 24 months as the target predictands. The model is trained using a rolling prediction strategy, with the RMSE of the Niño 3.4 index serving as the loss function to quantify the deviation between the predictions and the target values.

$$Loss =  \frac{1}{T_{out}} \sum_{t=1}^{T_{out}} \sqrt{(Nino 3.4_{t}^{out}-Nino 3.4_{t}^{tg})^2} $$
where $Nino 3.4_{t}^{out}$ and $Nino 3.4_{t}^{tg}$ represent the output and target Niño 3.4 index, respectively,  which are derived from normalized sea surface temperature anomalies at a depth of 5 meters. An Adam optimization algorithm is employed to optimize CTEFNet during training, with a learning rate warm-up technique applied, starting with an initial learning rate of $10^{-5}$.

\subsection{Gradient-based sensitivity analysis}

To reveal the ENSO signals extracted from CTEFNet across global patterns during prediction, we apply backpropagation to compute the gradient of each input point. The gradient quantifies the contribution of each input to the changes in the predicted Niño 3.4 index, allowing us to trace the precursors of ENSO events as a function of spatial location and lead time.

Specifically, we use the Niño 3.4 index from November of the ENSO years and subsequent months (when the index exceeds 0.5) within the valid period (1980-2021) as target values for prediction. Gradients are computed for each target value with respect to the corresponding inputs. The average gradient values across all target months are then used to determine the overall contribution of the input variables. This process is mathematically expressed as follows:

$$Grad_t= \frac{\partial Nino3.4_t} {\partial Inputs}\big|_{Inputs_t} $$
$$AvgGrad = mean_{t}\left(Grad_t\right) $$
where $Nino3.4_t$ and $Inputs_t$ represent the target predicted Niño 3.4 index and the input variables at target month $t$, respectively, $Grad_t$ denotes the gradient value of the input variables obtained at month $t$ through backpropagation in CTEFNet, and $AvgGrad$ represents the averaged gradients across all target months, reflecting the overall contribution of input variables to the Niño 3.4 index.

%% file: main_parts/data_availability.tex
\section*{Data Availability}

The data sources are listed below: CMIP6: \href{https://esgf-node.llnl.gov/ projects/cmip6/}{https://esgf-node.llnl.gov/ projects/cmip6/}; ORAS5: \href{https://cds.climate.copernicus.eu/datasets/reanalysis-oras5?tab=overview}{https://cds.climate.copernicus.eu/datasets/reanalysis-oras5?tab=overview}; ERA5: \href{https://cds.climate.copernicus.eu/datasets/reanalysis-era5-single-levels-monthly-means?tab=download}{https://cds.climate.copernicus.eu/datasets/reanalysis-era5-single-levels-monthly-means?tab=download}; GODAS: \href{https://psl.noaa.gov/data/gridded/data.godas.html}{https://psl.noaa.gov/data/gridded/data.godas.html}.

%% file: main_parts/contributions_interests.tex
\section*{Acknowledgements}

The research reported in this manuscript was supported by the National Natural Science Foundation of China (42276029).
 
 \section*{Author Contributions}

P. Zhan and Z.L. Wang are joint corresponding authors and contributed equally to conceptualization, project administration, funding acquisition, supervision, and writing-review and editing. Q. Chen and Y.H. Cui are joint first authors. Q. Chen contributed to data collection, model development, formal analysis, writing-original draft, and writing-review and editing. Y.H. Cui contributed to data processing, model development, investigation, and writing-original draft. G.B. Hong and Y.C. Pu provided technical support during the model training processes. K. Ashok, X.G. Zheng, X.Z. Zhang and W. Zhong reviewed the manuscript critically for important intellectual content. All authors discussed the results and commented on the manuscript.

\section*{Competing Interests}

The authors declare no competing interests.